\documentclass[10pt,journal]{IEEEtran}
\IEEEoverridecommandlockouts
\usepackage{array}
\usepackage[caption=false,font=normalsize,labelfont=sf,textfont=sf]{subfig}
\usepackage{textcomp}
\usepackage{stfloats}
\usepackage{url}
\usepackage{verbatim}
\usepackage{graphicx}
\usepackage{cite}
\usepackage{bm}
\usepackage{bbm}
\usepackage{color,soul}
\usepackage{multirow}
\usepackage{booktabs}
\usepackage{tabularx}
\usepackage[table,xcdraw]{xcolor}
\usepackage{enumitem}
\usepackage{verbatim}
\usepackage{mathrsfs} 
\usepackage{xcolor}
\usepackage{setspace}
\usepackage{hyperref}
\usepackage{algorithmicx}
\usepackage{float}
\usepackage{amsmath,amsthm,amsfonts,amssymb}
\theoremstyle{plain}
\newtheorem{thm}{Theorem}

\newtheorem{prop}[thm]{Proposition}

\newtheorem*{rem}{Remark}
\newtheorem{sty1}{Theorem}
\newtheorem{defi}[sty1]{Definition}

\usepackage[english]{babel}
\usepackage{algorithm}
\usepackage[noend]{algpseudocode}

\usepackage{tikz}
\usetikzlibrary{arrows.meta}

\allowdisplaybreaks[4]

\begin{document}

\title{Network-Native DNN Architecture Synthesis for Distributed Inference}

\author{Fan Yang, Xiaolong Wu, Qi Cao, Fen Hou, Yulin Shao
\thanks{F. Yang and F. Hou are with the State Key Laboratory of Internet of Things for Smart City, University of Macau, Macao, China.}
\thanks{X. Wu and Y. Shao are with the Department of Electrical and Computer Engineering, The University of Hong Kong, Hong Kong, China.}
\thanks{Qi Cao is with Xidian-Guangzhou Research Institute, Xidian University, Guangzhou, China.}
}

\maketitle

\begin{abstract}
Distributed deep neural network (DNN) inference typically follows a model-first paradigm, in which a predefined DNN is partitioned and deployed over a physical network. 
On irregular communication-computation topologies, however, this approach can create transmission bottlenecks and leave distributed resources underutilized. 
This paper reverses this perspective and asks: given a physical network, what kind of DNN can it naturally support?
To answer this question, we propose a network-native DNN architecture synthesis paradigm in which the physical network topology directly defines inter-node neural dependencies.
Neural computation is allocated across nodes under communication, computation, storage, synchronization, and latency constraints. 
We establish structural properties of the synthesis problem and develop linear programming (LP)-based Architecture Synthesis (LPAS), an efficient topology-guided algorithm that produces a feasible architecture with a computable bound on suboptimality. 
Experiments on physical networks of different sizes show that LPAS achieves relative gaps below 0.6\% to the LP-derived upper bound, with substantially lower runtime than direct mixed-integer LP optimization on larger networks.
Under matched network deadlines, the network-native architectures activate more compute nodes and utilize a larger proportion of available model-storage capacity than the model-first baseline.
After training, the network-native models achieve higher classification accuracy across all evaluated dataset-precision settings, with the average gain of up to 10.69\%.
This network-native design paradigm thus turns the physical network from a deployment constraint into an architecture design primitive.
\end{abstract}


\section{Introduction}
\label{sec:introduction}

The growing availability of computing resources across end devices, edge servers, access points, and programmable network devices \cite{ren2023survey,shao2021federated} creates new opportunities for distributed deep neural network (DNN) inference \cite{hao2026dnn}. By executing a DNN collaboratively across multiple networked nodes, distributed inference can pool fragmented computation and storage resources, reduce the burden on individual devices, and avoid dependence on a single execution point \cite{he2026accelerating}.

Against this backdrop, a natural question has dominated much of the existing literature: \emph{Given a DNN, how should it be partitioned and deployed onto a physical network?} This question has motivated a broad range of research on DNN partitioning, collaborative inference, split computing, and in-network AI. The prevailing deployment scenarios include two-tier device-cloud systems \cite{kang2017neurosurgeon,eshratifar2019jointdnn}, multi-node edge execution and pipeline parallelism \cite{he2020joint,hu2022pipeedge}, and hierarchical or graph-structured execution \cite{teerapittayanon2017distributed,hu2022distributed,huang2026partitioning}. These approaches generally specify the neural architecture first and then optimize its partitioning, placement, and execution over the network. We refer to this design pattern as the \emph{model-first deployment paradigm}.

On irregular communication-computation topologies, a fixed neural dependency graph can remain restrictive even after deployment is optimized. Placement and routing decisions determine where prescribed computations execute and how their features are transmitted, but generally preserve the DNN's dependencies. These dependencies can concentrate execution on a limited set of nodes and links while leaving resources on other branches underutilized. We refer to this incompatibility between the neural computational graph and the physical network as \emph{topological mismatch}.

In such a setting, the physical network is not merely a passive container for a pre-designed model. Its topology, computing capabilities, link characteristics, and synchronization structure actively constrain and shape the forms of neural computation that can be executed efficiently. This observation motivates us to ask the inverse question:

\begin{quote}
\emph{Given a physical network, what kind of DNN can it naturally support?}
\end{quote}

\begin{figure*}[t]
    \centering
    \includegraphics[width=0.85\textwidth]{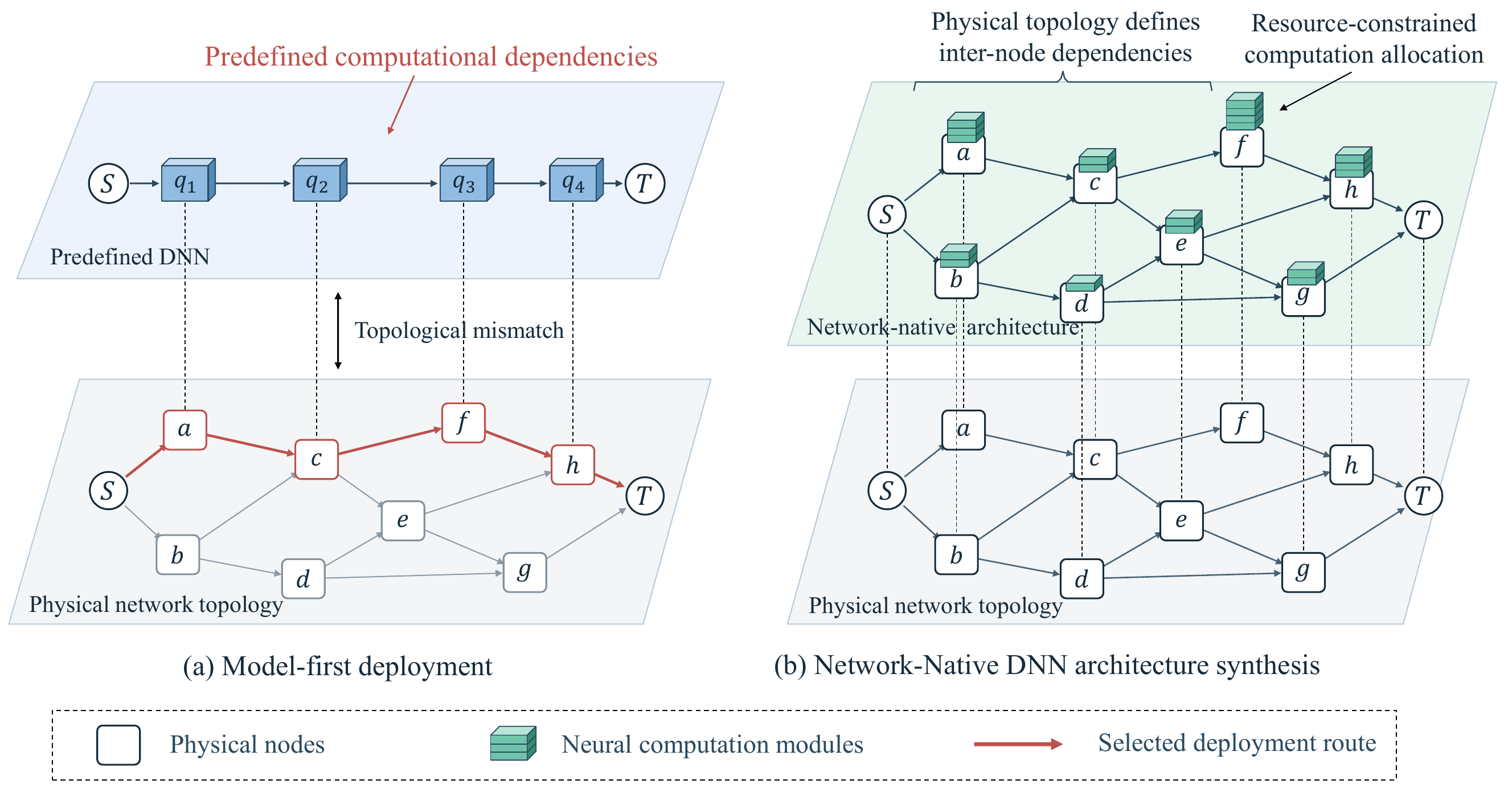}
    \caption{Comparison between model-first deployment and network-native DNN architecture synthesis. (a) A predefined DNN is mapped onto a selected route, illustrating topological mismatch through unused physical nodes and branches. (b) The synthesized architecture follows the physical topology in its inter-node neural dependencies, while network resources constrain the allocation of neural computation.}
    \label{fig:paradigm_comparison}
\end{figure*}

We address this question by proposing \emph{network-native DNN architecture synthesis}.
Rather than first designing a DNN and then searching for a deployment strategy, this approach starts from the physical communication-computation network and constructs a neural architecture aligned with it. We model the physical network as a directed acyclic graph (DAG), whose topology defines the inter-node neural dependencies of the synthesized architecture.
Node computation speeds, link delays, and storage capacities further determine the placement and amount of neural computation. The resulting model is therefore not a conventional DNN partitioned across multiple devices, but a \emph{network-native architecture} synthesized for the underlying communication-computation substrate. Fig.~\ref{fig:paradigm_comparison} illustrates the distinction between the two paradigms.

We synthesize architectures for a single inference request under a prescribed network-execution deadline and subsequently train them end-to-end for the target task. Our main contributions are summarized as follows.

\begin{itemize}[leftmargin=0.4cm]

\item We propose a network-native DNN architecture synthesis paradigm that fundamentally departs from conventional model-first distributed inference. Instead of partitioning and deploying a predefined DNN over a physical network, we construct a DNN whose inter-node neural dependencies are directly determined by the physical communication-computation topology. By treating the network as a directed acyclic graph (DAG), we formulate a resource-constrained architecture synthesis problem that maximizes the number of trainable neural modules, as a proxy for model capacity, while accounting for computation, communication, storage, synchronization, and end-to-end latency constraints.

\item We establish structural properties of the architecture synthesis problem that enable efficient optimization with theoretical guarantees. In particular, we characterize inference latency through longest paths on the physical DAG and derive an liner programming (LP)-based upper bound on the optimal number of neural modules. Building on these results, we develop LP-based architecture synthesis (LPAS), a polynomial-time algorithm that combines feasibility-preserving rounding with latency-aware greedy module allocation. LPAS guarantees a feasible architecture and provides a computable bound on its objective suboptimality, without requiring exact mixed-integer optimization.

\item We demonstrate the computational efficiency and practical benefits of network-native architecture synthesis through extensive evaluations across different physical network topologies and DNN configurations. LPAS achieves mean relative gaps below 0.6\% to the LP-derived upper bound while substantially reducing runtime compared with direct mixed-integer linear programming (MILP) optimization on larger networks. Under matched network-execution deadlines, the synthesized architectures exploit more distributed computing and storage resources than an optimized model-first deployment baseline. Experiments on CIFAR-10 and CIFAR-100 further show higher classification accuracy across all evaluated dataset-precision settings, with gains of up to 10.69\%.

\end{itemize}

\section{Related Work}
\label{sec:related_work}

Distributed DNN inference has been extensively studied under the model-first paradigm \cite{ren2023survey,hao2026dnn}. Early device-cloud systems, such as Neurosurgeon \cite{kang2017neurosurgeon} and JointDNN \cite{eshratifar2019jointdnn}, determine how a predefined DNN should be partitioned between two endpoints based on execution profiles, latency, and energy considerations.

As distributed inference extends to multiple devices, deployment decisions must also account for computation placement and data-transfer paths \cite{bao2025joint,shao2024theory}. This has led to joint optimization of DNN partition deployment and computing-resource allocation in MEC networks \cite{he2020joint}, as well as layered-graph formulations that capture computation-node and data-transfer-path selection \cite{jung2023optimization}. For heterogeneous edge devices, pipeline-based approaches assign predefined stages to devices \cite{hu2022pipeedge} or jointly optimize device-to-stage assignment and inter- and intra-layer partitioning to improve inference throughput \cite{dai2025joint}.

More complex distributed inference settings involve hierarchical collaboration and DNNs with nonsequential computational dependencies \cite{singhal2024resource}. Hierarchical and cooperative approaches distribute jointly trained model components across device-edge-cloud systems \cite{teerapittayanon2017distributed} or adaptively partition DNN workloads across heterogeneous edge devices \cite{zeng2020coedge}. When the DNN itself contains parallel branches and merges, deployment must also preserve dependencies among computations that may execute on different devices. For these DAG-structured DNNs, existing work addresses dependency-aware execution \cite{hu2022distributed}, equivalent chain representations for streaming partitioning and offloading \cite{wu2024pdd}, and joint operator-grained partitioning, offloading, and resource allocation in multi-user MEC systems \cite{zhang2026taskaware}.

Beyond deployment optimization, some studies modify DNN architectures to improve distributed execution. DFSNet transforms conventional DNNs into dividing-fuse architectures through channel-wise partitioning and sparse fusion, with fusion positions selected through search \cite{hou2022dfsnet}. Other approaches restructure, partition, and prune pretrained DNNs to reduce dependencies among sub-models \cite{abdi2023efficient}, or incorporate network-dependent communication costs into DNN pruning \cite{jian2023communication}.

Neural architecture search (NAS) provides another approach to DNN architecture design by exploring candidate structures within a predefined search space. Hardware-aware methods, such as MnasNet \cite{tan2019mnasnet} and ProxylessNAS \cite{cai2018proxylessnas}, incorporate device-dependent performance into the selection of operators, blocks, widths, and depths. When DNN inference extends across multiple devices, architecture design must also consider how neural computation is distributed and how intermediate features are transmitted. For example, LENS incorporates communication-dependent latency and energy into architecture search for edge-cloud systems \cite{odema2021lens}, while SplitNets jointly considers architecture design, model splitting, and communication reduction in sensor-aggregator systems \cite{dong2022splitnets}. The coupling between architecture search and deployment is further explored in different ways. JMSNAS searches candidate operations under task-loss and network-latency objectives with a predefined computational graph and device-wise layer partitions \cite{tian2022jmsnas}, whereas GCoDE jointly searches graph neural network architectures and operation mappings across device-edge systems \cite{zhou2024gcode}.

Existing studies range from deploying predefined DNNs to adapting or searching neural architectures under system constraints. In contrast, network-native DNN architecture synthesis uses the physical communication-computation DAG to directly define the inter-node neural dependencies of the synthesized architecture: physical branches and merges define neural branches and fusion points. This network-defined structure leads to a constrained module-allocation problem whose DAG properties support explicit latency analysis and efficient architecture synthesis with feasibility and gap guarantees.

\section{Problem Formulation}
\label{problem_formulation}

We consider the synthesis of a DNN architecture and its execution over a prescribed physical communication-computation network.
The physical network is modeled as a DAG \(\mathcal{G}=(\mathcal{V},\mathcal{E})\), where \(\mathcal{V}\) is the set of physical nodes and \(\mathcal{E}\) is the set of directed communication links.

A source node \(S\in\mathcal{V}\) receives the input sample, and a sink node \(T\in\mathcal{V}\) produces the final output. For a specific learning task, fixed input and output interfaces map the raw input to the common feature space and map the terminal feature or features to the task output, respectively. These interfaces are not architecture-synthesis variables. We therefore define the modeled execution interval to begin when the common input feature becomes available at \(S\) and to end when all features required by the output interface have arrived at \(T\).

Accordingly, \(S\) and \(T\) serve as the boundary nodes of the distributed architecture, while neural computation modules are assigned only to the intermediate nodes. Let \(\mathcal{I}=\mathcal{V}\setminus\{S,T\}\) denote the set of intermediate compute nodes, and let \(N=|\mathcal{I}|\).

We assume that every physical node lies on at least one directed path from \(S\) to \(T\); any node that violates this condition can be removed without affecting the modeled inference execution.

For every \(v\in\mathcal{V}\), let \(d_v=|\{u:(u,v)\in\mathcal{E}\}|\) denote its in-degree. Each intermediate node \(v\in\mathcal{I}\) is characterized by a computation-delay coefficient \(A_v>0\) and an available model-storage capacity \(M_v\geq0\). The coefficient \(A_v\) is measured in seconds per unit computation; hence, a smaller \(A_v\) corresponds to a faster compute node. Each directed link \((u,v)\in\mathcal{E}\) is characterized by a communication-delay coefficient \(B_{uv}>0\), measured in seconds per unit transmitted data. Equivalently, \(B_{uv}\) is the reciprocal of the effective transmission rate of the link. Consequently, transmitting a feature of data size \(D\) over link \((u,v)\) incurs a delay of \(B_{uv}D\). Different physical links are assumed to be independently provisioned, so simultaneous transmissions on distinct links do not contend for communication resources.

\subsection{Modular Architecture Design}
\label{primitive}

For a given task, the basic neural module, feature representation, and fusion operator are selected before synthesis. The physical DAG determines the inter-node computational dependencies, while node-wise module allocation determines the amount of neural computation. We adopt the following modular design assumptions.

\begin{enumerate}
\item All features transmitted over the edges of \(\mathcal{G}\) have the same data size \(D\).

\item An intermediate node begins its local processing only after all incoming features have arrived.

\item The trainable architecture on a node is constructed from repeated instances of a dimension-preserving neural module (see Definition~\ref{defi:module}).

\item Transmissions over distinct outgoing links proceed in parallel (in a multicast fashion).
\end{enumerate}

\begin{defi}[Module]
\label{defi:module}
A module is a neural computation unit that maps an input feature of size \(D\) to an output feature of the same size \(D\). It can be instantiated using a standard neural building block, such as a residual block \cite{he2016deep}, a self-attention block \cite{vaswani2017attention}, or a depthwise-separable convolution block \cite{fran2017deep}.
\end{defi}

For a fixed module primitive and feature shape, let \(F>0\) denote the computation required by one module. The value of \(F\) can be derived analytically from the selected neural operators or obtained through offline profiling. Once the module primitive and feature shape have been selected, \(F\) is fixed throughout the architecture synthesis problem.

Each intermediate physical node can host a sequential stack of repeated modules. For \(v\in\mathcal{I}\), let \(L_v\in\mathbb{Z}_{\geq0}\) denote the number of modules assigned to node \(v\). The complete module allocation is represented by \(\boldsymbol{L}=(L_v)_{v\in\mathcal{I}}\). For notational convenience, we extend the allocation to the boundary nodes using \(L_S=L_T=0\).

The local processing performed at an intermediate node consists of an optional fusion operation followed by its assigned module stack. If \(d_v=1\), the single incoming feature is passed directly to the local modules. If \(d_v>1\), the incoming features are first fused into one feature of size \(D\). The fusion operator is selected before synthesis rather than optimized during module allocation.

Let \(G_v\geq0\) denote the fixed computation cost of the prescribed fusion operation at node \(v\), with \(G_v=0\) when \(d_v=1\). For a merging node, \(G_v\) may depend on the feature shape, the in-degree \(d_v\), and the selected fusion implementation, but is independent of the module allocation \(L_v\).

The processing delay at an intermediate node \(v\in\mathcal{I}\) is therefore
\(\Delta_v(L_v) = A_v(FL_v+G_v)\).
For the boundary nodes, we use the convention
\(\Delta_S(L_S)=\Delta_T(L_T)=0\).

After completing its local processing, an intermediate node replicates its output feature and transmits one copy over each outgoing link. The replication itself introduces no additional neural computation, while the transmission delay is accounted for separately on each link.

The physical DAG remains fixed throughout synthesis. In particular, setting \(L_v=0\) removes only the repeated module stack at node \(v\); it does not deactivate the node or its incident links. Such a node performs the prescribed fusion operation when \(d_v>1\), or acts as an identity relay when \(d_v=1\), before forwarding its output to its child nodes.

\begin{figure*}[t]
    \centering
    \includegraphics[width=0.9\textwidth]{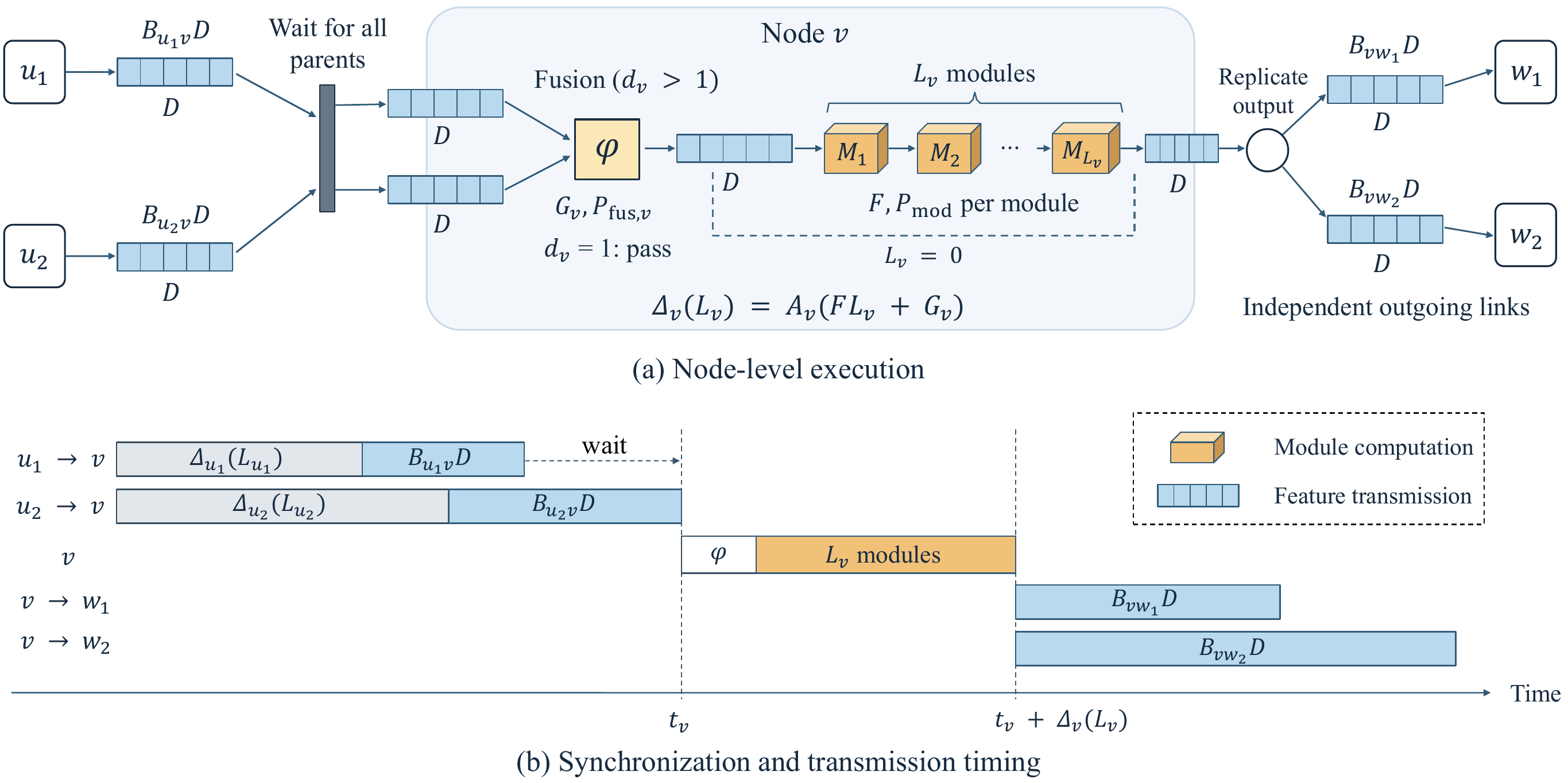}
    \caption{Execution at an intermediate node.
    (a) Node-level synchronization, fusion, module execution, and
    output replication.
    (b) Timing of incoming feature arrivals, node computation,
    and independent outgoing transmissions.}
    \label{fig:netdnn_node_execution}
\end{figure*}

\subsection{Resource and Latency Constraints}
\label{resource_latency_constraints}

\subsubsection{Node Storage Constraint}
\label{storage_constraint}

In practical distributed inference systems, each physical node has limited storage for neural model parameters. Therefore, a node cannot host an arbitrarily large number of modules even if its computation speed is high.

Let \(P_{\mathrm{mod}}>0\) denote the parameter storage required by one instance of the selected neural module. As with the per-module computation \(F\), \(P_{\mathrm{mod}}\) can be determined analytically or through offline profiling and remains fixed during synthesis.

Let \(P_{\mathrm{fus},v}\geq0\) denote the fixed parameter-storage cost of the prescribed fusion operation at node \(v\), with \(P_{\mathrm{fus},v}=0\) when \(d_v=1\).

The parameter-storage constraint at any intermediate node \(v\in\mathcal{I}\) is
\(
L_vP_{\mathrm{mod}}+P_{\mathrm{fus},v} \leq M_v.
\)

A feasible instance requires \(M_v\geq P_{\mathrm{fus},v}\) for every \(v\in\mathcal{I}\). Since \(L_v\) is integer, the storage constraint is equivalently represented by the module-capacity bound
\begin{equation}
0\leq L_v\leq U_v
\triangleq
\left\lfloor
\frac{M_v-P_{\mathrm{fus},v}}
{P_{\mathrm{mod}}}
\right\rfloor,
\qquad
\forall v\in\mathcal{I}.
\label{eq:capacity_bound}
\end{equation}
For notational completeness, we set \(U_S=U_T=0\). The same capacity bound is retained in the continuous relaxation. The storage model accounts only for the parameters of the deployed inference model; activation buffers, temporary operator workspace, communication buffers, and runtime memory are assumed to be provisioned separately.

\subsubsection{Edge-Level Latency Constraint}
\label{edge_latency_constraint}

The primary system-level requirement is an end-to-end latency deadline \(T_D\). For each intermediate node \(v\in\mathcal{I}\), let \(t_v\) denote the time at which its local processing begins. Consistent with the boundary convention introduced above, \(t_S\) is the time at which the common source feature becomes available, while \(t_T\) is the time at which all features required by the output interface have arrived at the sink.

For every directed edge \((u,v)\in\mathcal{E}\), the feature transmitted by node \(u\) cannot become available at node \(v\) before node \(u\) has completed its local computation and transmitted its output feature. Therefore, the following edge-level timing constraint must hold:
\begin{equation}
t_v \ge t_u+\Delta_u(L_u)+B_{uv}D,
\quad
\forall (u,v)\in\mathcal{E}.
\label{eq:edge_timing}
\end{equation}

The source node starts at \(t_S=0\). The sink node must be reached before the deadline: \(t_T\leq T_D\).

The edge-level formulation automatically captures synchronization at merging nodes. If a node \(v\) has multiple incoming edges, then \eqref{eq:edge_timing} must hold for every parent \(u\) of \(v\). As a result, \(t_v\) must be no smaller than the latest arrival time among all incoming features. Therefore, no additional synchronization constraint is required for a single inference request. The complete execution flow of an intermediate node is summarized in Fig.~\ref{fig:netdnn_node_execution}.

\subsection{The Architecture Synthesis Problem}
\label{main_problem}

We next specify the objective of architecture synthesis. Prior studies on deep residual networks and CNN scaling have shown that increasing model depth and size can improve task performance under suitable architectural and training conditions \cite{he2016deep,tan2019efficientnet}. Research on neural scaling laws further highlights the relationship between model scale and learning performance \cite{bahri2024explaining}. These findings motivate a capacity-oriented design that exploits the available network resources to construct larger trainable models.

In this modular design, each module contributes the same computation \(F\) and parameter storage \(P_{\mathrm{mod}}\), while fusion costs are fixed independently of the module allocation. Thus, the total number of modules directly determines the aggregate computation and parameter count of the trainable module stacks. We therefore use the network-wide module budget
\(\mathcal{Z}(\boldsymbol{L})=\sum_{v\in\mathcal{I}}L_v\)
as a tractable proxy for the representational capacity of the synthesized architecture and maximize it under the physical-network constraints. The task performance of the resulting architectures is subsequently evaluated after training.

The architecture synthesis problem is therefore to determine how many modules should be instantiated at each intermediate node while satisfying the node-capacity and end-to-end latency requirements. Formally, we seek a module allocation \(\boldsymbol{L}\) and a feasible set of node start times \(\{t_v\}_{v\in\mathcal{V}}\) according to
\begin{equation}
\label{eq:netdnn_problem}
\begin{aligned}
\max_{\boldsymbol{L},\,\{t_v\}_{v\in\mathcal{V}}}\quad & \mathcal{Z}(\boldsymbol{L})\\
\mathrm{s.t.}\quad & t_v\geq t_u+\Delta_u(L_u)+B_{uv}D, && \forall (u,v)\in\mathcal{E},\\
& t_S=0,\qquad t_T\leq T_D,\\
& 0\leq L_v\leq U_v, \quad L_v\in\mathbb{Z}_{\geq0}, && \forall v\in\mathcal{I}.
\end{aligned}
\end{equation}

The physical DAG fixes the inter-node connectivity, whereas \(\boldsymbol{L}\) determines the trainable capacity placed at each intermediate node. The constraints capture computation, communication, synchronization, and storage dependencies. Because \(\Delta_v(L_v)\) is affine in \(L_v\) and the allocation variables are integer-valued, the design formulation in \eqref{eq:netdnn_problem} admits a MILP representation. This representation provides a computational means of synthesizing DNN architectures; the underlying architectural question is how the trainable neural capacity should be distributed over a given physical communication-computation network.

\section{Structural Properties of the Synthesis Problem}
\label{structural_analysis}

The design of LPAS relies on two key ingredients: a bound on the optimal module budget and efficient methods for constructing and evaluating feasible allocations. We first establish an LP-derived upper bound and characterize allocation feasibility through longest-path recursions on the physical DAG. These results then enable a feasible floor-rounded initialization and an exact test for one-module extensions.

\subsection{Continuous Relaxation and LP Upper Bound}
\label{lp_relaxation}

Removing the integrality requirement from \eqref{eq:netdnn_problem} while retaining all other constraints gives the continuous relaxation
\begin{equation}
\label{eq:lp_relaxation}
\begin{aligned}
\max_{\boldsymbol{L},\,\{t_v\}_{v\in\mathcal{V}}}\quad & \mathcal{Z}(\boldsymbol{L})\\
\mathrm{s.t.}\quad & t_v\geq t_u+\Delta_u(L_u)+B_{uv}D, && \forall (u,v)\in\mathcal{E},\\
& t_S=0,\qquad t_T\leq T_D,\\
& 0\leq L_v\leq U_v, && \forall v\in\mathcal{I}.
\end{aligned}
\end{equation}

Since \(\Delta_v(L_v)\) is affine in \(L_v\), \eqref{eq:lp_relaxation} is a linear program. Whenever it is feasible, let \(\boldsymbol{L}^{\mathrm{LP}}\) denote an optimal allocation and define \(\mathcal{Z}_{\mathrm{LP}}^\star=\mathcal{Z}(\boldsymbol{L}^{\mathrm{LP}})\). Let \(\mathcal{Z}_{\mathrm{MILP}}^\star\) denote the optimal objective value of \eqref{eq:netdnn_problem} whenever the integer problem is feasible.

Every integer-feasible solution is also feasible for \eqref{eq:lp_relaxation}. Since \(\mathcal{Z}_{\mathrm{MILP}}^\star\) is integer-valued, it follows that
\begin{equation}
\mathcal{Z}_{\mathrm{MILP}}^\star\leq\mathcal{Z}_{\mathrm{UB}}\triangleq\left\lfloor\mathcal{Z}_{\mathrm{LP}}^\star\right\rfloor\leq\mathcal{Z}_{\mathrm{LP}}^\star.
\label{eq:integer_upper_bound}
\end{equation}

Thus, infeasibility of \eqref{eq:lp_relaxation} implies infeasibility of the integer problem, while \(\mathcal{Z}_{\mathrm{UB}}\) is a valid integer upper bound for every feasible instance.

\subsection{Longest-Path Latency Characterization}
\label{longest_path_eval}

Consider a fixed allocation \(\boldsymbol{L}\in\mathbb{R}_{\geq0}^{|\mathcal{I}|}\), with \(L_S=L_T=0\). For each edge \((u,v)\in\mathcal{E}\), define the allocation-dependent weight \(w_{uv}(\boldsymbol{L})=\Delta_u(L_u)+B_{uv}D\), which includes the processing delay at node \(u\) and the transmission delay over \((u,v)\).

Let \(\theta_S^{\boldsymbol{L}}(v)\) denote the maximum latency from \(S\) to the start of node \(v\), with \(\theta_S^{\boldsymbol{L}}(S)=0\). Since \(\mathcal{G}\) is acyclic, the remaining values can be computed in topological order as
\begin{equation}
\theta_S^{\boldsymbol{L}}(v)=\max_{(u,v)\in\mathcal{E}}\left\{\theta_S^{\boldsymbol{L}}(u)+w_{uv}(\boldsymbol{L})\right\},\qquad v\neq S.
\label{eq:forward_dp}
\end{equation}
The earliest achievable end-to-end latency under \(\boldsymbol{L}\) is \(T(\boldsymbol{L})=\theta_S^{\boldsymbol{L}}(T)\).

For the one-module extension test, let \(\theta_T^{\boldsymbol{L}}(v)\) denote the maximum remaining latency from the start of node \(v\) to the start of \(T\), including the processing delay at \(v\), with \(\theta_T^{\boldsymbol{L}}(T)=0\). The remaining values satisfy
\begin{equation}
\theta_T^{\boldsymbol{L}}(v)=\max_{(v,q)\in\mathcal{E}}\left\{w_{vq}(\boldsymbol{L})+\theta_T^{\boldsymbol{L}}(q)\right\},\qquad v\neq T.
\label{eq:backward_dp}
\end{equation}
Both recursions can be evaluated with one traversal in each direction of the topological order.

\begin{prop}[Fixed-Allocation Feasibility]
\label{prop:allocation_feasibility}
For any fixed allocation \(\boldsymbol{L}\in\mathbb{R}_{\geq0}^{|\mathcal{I}|}\), there exist timing variables satisfying all edge-level timing constraints, \(t_S=0\), and \(t_T\leq T_D\) if and only if
\begin{equation}
T(\boldsymbol{L})\leq T_D.
\label{eq:latency_feasibility}
\end{equation}
Consequently, \(\boldsymbol{L}\) is feasible for \eqref{eq:lp_relaxation} if and only if \eqref{eq:latency_feasibility} holds and \(0\leq L_v\leq U_v\) for every \(v\in\mathcal{I}\). If \(\boldsymbol{L}\) is integer-valued, the same conditions characterize feasibility in \eqref{eq:netdnn_problem}.
\end{prop}

\begin{proof}
Consider any timing variables satisfying the edge-level constraints and \(t_S=0\). Induction along a topological order gives \(t_v\geq\theta_S^{\boldsymbol{L}}(v)\) for every \(v\in\mathcal{V}\). Hence, \(t_T\leq T_D\) requires \(T(\boldsymbol{L})=\theta_S^{\boldsymbol{L}}(T)\leq T_D\).

Conversely, suppose that \(T(\boldsymbol{L})\leq T_D\) and set \(t_v=\theta_S^{\boldsymbol{L}}(v)\) for every \(v\in\mathcal{V}\). The recursion in \eqref{eq:forward_dp} ensures that every edge-level timing constraint is satisfied, while \(t_S=0\) and \(t_T=T(\boldsymbol{L})\leq T_D\). This proves the equivalence.
\end{proof}

Thus, \(\theta_S^{\boldsymbol{L}}(v)\) is the earliest feasible start time of node \(v\). A source-to-sink path whose weight equals \(T(\boldsymbol{L})\) is called a critical path under \(\boldsymbol{L}\).

\subsection{Floor-Rounded Integer Initialization}
\label{monotonicity_floor}

Assume that \eqref{eq:lp_relaxation} is feasible. Define the componentwise floor-rounded allocation as
\begin{equation}
\boldsymbol{L}^0=\left\lfloor\boldsymbol{L}^{\mathrm{LP}}\right\rfloor\triangleq\left(\left\lfloor L_v^{\mathrm{LP}}\right\rfloor\right)_{v\in\mathcal{I}},
\label{eq:floor_allocation}
\end{equation}
where the floor operation is applied componentwise, and let \(\mathcal{Z}_0=\mathcal{Z}(\boldsymbol{L}^0)\).

\begin{prop}[Floor-Rounding Guarantee]
\label{prop:floor_feasibility}
The floor-rounded allocation \(\boldsymbol{L}^0\) is feasible for the integer problem in \eqref{eq:netdnn_problem}. Consequently, the integer optimum \(\mathcal{Z}_{\mathrm{MILP}}^\star\) is well defined and, if \(N=|\mathcal{I}|\geq1\),
\begin{equation}
0\leq\mathcal{Z}_{\mathrm{MILP}}^\star-\mathcal{Z}_0\leq\mathcal{Z}_{\mathrm{UB}}-\mathcal{Z}_0\leq N-1.
\label{eq:floor_additive_gap}
\end{equation}
\end{prop}

\begin{proof}
By construction, \(L_v^0\leq L_v^{\mathrm{LP}}\) for every \(v\in\mathcal{I}\). Since \(\Delta_v(L_v)\) is nondecreasing in \(L_v\), every edge weight and therefore every source-to-sink path latency under \(\boldsymbol{L}^0\) is no larger than under \(\boldsymbol{L}^{\mathrm{LP}}\). Proposition~\ref{prop:allocation_feasibility} then gives
\(
T(\boldsymbol{L}^0)\leq T(\boldsymbol{L}^{\mathrm{LP}})\leq T_D.
\)
Moreover, \(0\leq L_v^{\mathrm{LP}}\leq U_v\) and \(U_v\) is integer, so \(0\leq L_v^0\leq U_v\). Hence, \(\boldsymbol{L}^0\) is integer-feasible.

For the objective bound, let \(r_v=L_v^{\mathrm{LP}}-L_v^0\), where \(0\leq r_v<1\). Since \(\mathcal{Z}_0\) is integer,
\begin{equation}
\label{eq:rounding_gap_step}
Z_{\mathrm{UB}}-Z_0=\left\lfloor\sum_{v\in\mathcal I}r_v\right\rfloor\le N-1.
\end{equation}
Since \(\boldsymbol{L}^0\) is feasible, \(Z_0\le Z_{\mathrm{MILP}}^\star\), while \eqref{eq:integer_upper_bound} gives \(Z_{\mathrm{MILP}}^\star\le Z_{\mathrm{UB}}\). Combining these inequalities with \eqref{eq:rounding_gap_step} completes the proof.
\end{proof}

Together with the inclusion of the integer feasible set in the LP feasible set, Proposition~\ref{prop:floor_feasibility} also shows that the LP relaxation is feasible if and only if the integer problem is feasible.

\subsection{Exact One-Module Extension}
\label{one_module_extension}

Let \(\boldsymbol{L}\) be a feasible integer allocation. For \(v\in\mathcal{I}\), let \(\mathbf{e}_v\) denote the corresponding unit vector and define
\begin{equation}
\boldsymbol{L}^{(v)}\triangleq\boldsymbol{L}+\mathbf{e}_v.
\label{eq:one_module_extension}
\end{equation}

\begin{prop}[Exact One-Module Extension Test]
\label{prop:one_module_extension}
The allocation \(\boldsymbol{L}^{(v)}\) is feasible if and only if
\begin{gather}
L_v+1\leq U_v,
\label{eq:extension_capacity_test}\\
\theta_S^{\boldsymbol{L}}(v)+\theta_T^{\boldsymbol{L}}(v)+A_vF\leq T_D.
\label{eq:extension_latency_test}
\end{gather}
\end{prop}

\begin{proof}
Condition~\eqref{eq:extension_capacity_test} is necessary and sufficient for the node-capacity constraint because only \(L_v\) changes. Adding one module to node \(v\) increases every outgoing edge weight of \(v\) by \(A_vF\), while all other edge weights remain unchanged.

The maximum latency of a source-to-sink path passing through \(v\) is \(\theta_S^{\boldsymbol{L}}(v)+\theta_T^{\boldsymbol{L}}(v)\). Since \(\mathcal{G}\) is acyclic, the maximizing source-to-\(v\) path and \(v\)-to-sink path cannot share any node other than \(v\). They can therefore be concatenated into a valid source-to-sink path. Therefore,
\begin{equation}
\label{eq:module_extension_latency}
T(\boldsymbol{L}^{(v)})=\max\bigl\{T(\boldsymbol{L}),\theta_S^{\boldsymbol{L}}(v)+\theta_T^{\boldsymbol{L}}(v)+A_vF\bigr\}.
\end{equation}
Because $\boldsymbol{L}$ is feasible, \(T(\boldsymbol{L})\leq T_D\). Therefore, by \eqref{eq:module_extension_latency}, the extended allocation satisfies the latency deadline if and only if \eqref{eq:extension_latency_test} holds.
\end{proof}

Proposition~\ref{prop:one_module_extension} tests each one-module extension without enumerating source-to-sink paths. For a fixed feasible allocation, one forward and one reverse topological traversal suffice to identify all feasible one-module extensions in \(O(|\mathcal{V}|+|\mathcal{E}|)\) time.

\section{Efficient LP-Based Architecture Synthesis}
\label{lpas}

Building on the structural properties established in Section~\ref{structural_analysis}, we develop an efficient synthesis method called \emph{LP-based Architecture Synthesis} (LPAS). The method first solves the continuous relaxation and rounds its solution down to obtain a feasible integer module allocation. Starting from this allocation, LPAS repeatedly adds one module at a time using the exact feasibility test in Proposition~\ref{prop:one_module_extension}. Each iteration selects the feasible extension with the smallest resulting end-to-end latency.

\subsection{Initialization}
\label{lpas_initialization}

LPAS begins by solving the continuous relaxation in \eqref{eq:lp_relaxation}. If the relaxation is infeasible, the integer problem is also infeasible, and the method terminates.

Otherwise, let \(\boldsymbol{L}^{\mathrm{LP}}\) denote an optimal relaxed allocation, and recall the integer upper bound \(\mathcal{Z}_{\mathrm{UB}}=\lfloor\mathcal{Z}_{\mathrm{LP}}^\star\rfloor\). The initial integer allocation is obtained by componentwise floor rounding as \(\boldsymbol{L}^{0}=\lfloor\boldsymbol{L}^{\mathrm{LP}}\rfloor\), with \(\mathcal{Z}_0=\mathcal{Z}(\boldsymbol{L}^{0})\). By Proposition~\ref{prop:floor_feasibility}, \(\boldsymbol{L}^{0}\) is feasible for the integer problem.

We define \(K_{\max}\triangleq\mathcal{Z}_{\mathrm{UB}}-\mathcal{Z}_0\). By \eqref{eq:floor_additive_gap}, \(0\leq K_{\max}\leq N-1\).

If the LP has multiple optimal solutions, any optimal allocation may be used. The resulting initialization and architecture may vary, but the guarantees below hold for every such choice. Given an initialization, a fixed node ordering makes the padding deterministic.

\subsection{Greedy Padding}
\label{lpas_module_addition}

For a current feasible integer allocation \(\boldsymbol{L}\), let \(\mathcal{C}(\boldsymbol{L})\subseteq\mathcal{I}\) contain the nodes satisfying both the capacity test in \eqref{eq:extension_capacity_test} and the latency test in \eqref{eq:extension_latency_test}. Proposition~\ref{prop:one_module_extension} makes this exactly the set of feasible one-module extensions.

The resulting end-to-end latency $T(\boldsymbol{L}^{(v)})$ of each candidate extension can be evaluated exactly using \eqref{eq:module_extension_latency}.

Starting from the allocation \(\boldsymbol{L}^{0}\), LPAS selects \(v^\star\in\arg\min_{v\in\mathcal{C}(\boldsymbol{L})}T(\boldsymbol{L}^{(v)})\), resolves ties by the fixed node ordering, and updates \(\boldsymbol{L}\leftarrow\boldsymbol{L}+\mathbf{e}_{v^\star}\). The forward and reverse longest-path values are recomputed at the start of each subsequent padding iteration. The procedure stops when \(\mathcal{C}(\boldsymbol{L})=\emptyset\) or the module budget reaches \(\mathcal{Z}_{\mathrm{UB}}\), returning \(\hat{\boldsymbol{L}}\).

\begin{rem}[Beam Search]
The padding stage can be extended to a beam-search variant by retaining up to \(W>1\) feasible candidate allocations at each iteration instead of a single one. Each retained allocation generates its feasible one-module extensions, from which up to \(W\) candidates with the smallest resulting end-to-end latencies are selected for the next iteration. When \(W=1\), this procedure reduces to the greedy padding strategy described above. We adopt \(W=1\) in this work to avoid maintaining and evaluating multiple candidate allocations, thereby keeping the algorithm simple and computationally efficient.
\end{rem}

\subsection{Algorithm, Guarantees, and Complexity}
\label{lpas_algorithm}

For the returned allocation, define the LP-induced upper-bound gap as \(G_{\mathrm{UB}}\triangleq\mathcal{Z}_{\mathrm{UB}}-\mathcal{Z}(\hat{\boldsymbol{L}})\). Algorithm~\ref{alg:lpas} summarizes the complete procedure.

\begin{algorithm}[t]
\caption{LPAS: LP-Based Architecture Synthesis}
\label{alg:lpas}
\begin{algorithmic}[1]
\State{\textbf{Input:} \(\mathcal{G}=(\mathcal{V},\mathcal{E})\), \(\{A_v\}\), \(\{B_{uv}\}\), \(D\), \(F\), \(\{G_v\}\), \(\{U_v\}\), \(T_D\)}
\State{\textbf{Output:} \(\hat{\boldsymbol{L}}\), \(\mathcal{Z}(\hat{\boldsymbol{L}})\), \(\mathcal{Z}_{\mathrm{UB}}\), \(G_{\mathrm{UB}}\), \(T(\hat{\boldsymbol{L}})\)}
\State Solve the LP relaxation in \eqref{eq:lp_relaxation}
\If{the LP relaxation is infeasible}
    \State \Return infeasible
\EndIf
\State Obtain \(\boldsymbol{L}^{\mathrm{LP}}\) and compute \(\mathcal{Z}_{\mathrm{UB}}\) from \eqref{eq:integer_upper_bound}
\State \(\hat{\boldsymbol{L}}\leftarrow\lfloor\boldsymbol{L}^{\mathrm{LP}}\rfloor\)
\While{\(\mathcal{Z}(\hat{\boldsymbol{L}})<\mathcal{Z}_{\mathrm{UB}}\)}
    \State Compute \(\theta_S^{\hat{\boldsymbol{L}}}\) and \(\theta_T^{\hat{\boldsymbol{L}}}\)
    \State Construct \(\mathcal{C}(\hat{\boldsymbol{L}})\)
    \If{\(\mathcal{C}(\hat{\boldsymbol{L}})=\emptyset\)}
        \State \textbf{break}
    \EndIf
    \State Compute \(T(\hat{\boldsymbol{L}}^{(v)})\) for all \(v\in\mathcal{C}(\hat{\boldsymbol{L}})\)
    \State Select \(v^\star\) with minimum \(T(\hat{\boldsymbol{L}}^{(v)})\), breaking ties by the fixed node ordering
    \State \(\hat{\boldsymbol{L}}\leftarrow\hat{\boldsymbol{L}}+\mathbf{e}_{v^\star}\)
\EndWhile
\State Compute \(T(\hat{\boldsymbol{L}})\) using the forward longest-path recursion
\State \(G_{\mathrm{UB}}\leftarrow\mathcal{Z}_{\mathrm{UB}}-\mathcal{Z}(\hat{\boldsymbol{L}})\)
\State \Return \(\hat{\boldsymbol{L}}\), \(\mathcal{Z}(\hat{\boldsymbol{L}})\), \(\mathcal{Z}_{\mathrm{UB}}\), \(G_{\mathrm{UB}}\), \(T(\hat{\boldsymbol{L}})\)
\end{algorithmic}
\end{algorithm}

\begin{prop}[LPAS Guarantees]
\label{prop:lpas_guarantees}
Suppose that the LP relaxation is feasible and \(N\geq1\). The allocation \(\hat{\boldsymbol{L}}\) returned by LPAS is feasible for \eqref{eq:netdnn_problem} and satisfies
\begin{equation}
0\leq\mathcal{Z}_{\mathrm{MILP}}^\star-\mathcal{Z}(\hat{\boldsymbol{L}})\leq G_{\mathrm{UB}}\leq K_{\max}\leq N-1.
\label{eq:lpas_gap_guarantee}
\end{equation}
\end{prop}

\begin{proof}
The initial allocation \(\boldsymbol{L}^0\) is feasible by Proposition~\ref{prop:floor_feasibility}. Every subsequent allocation is obtained using a feasible one-module extension identified by Proposition~\ref{prop:one_module_extension}, so \(\hat{\boldsymbol{L}}\) is feasible. Hence, \(\mathcal{Z}(\hat{\boldsymbol{L}})\leq\mathcal{Z}_{\mathrm{MILP}}^\star\).

Since LPAS only inserts modules into \(\boldsymbol{L}^0\), we have \(\mathcal{Z}_0\leq\mathcal{Z}(\hat{\boldsymbol{L}})\leq\mathcal{Z}_{\mathrm{UB}}\). Together with \(\mathcal{Z}_{\mathrm{MILP}}^\star\leq\mathcal{Z}_{\mathrm{UB}}\), this gives
\begin{equation}
\label{eq:lpas_gap_step}
0\leq\mathcal{Z}_{\mathrm{MILP}}^\star-\mathcal{Z}(\hat{\boldsymbol{L}})\leq G_{\mathrm{UB}}\leq\mathcal{Z}_{\mathrm{UB}}-\mathcal{Z}_0=K_{\max}.
\end{equation}
Finally, combining \eqref{eq:lpas_gap_step} with the bound \(K_{\max}\leq N-1\) from \eqref{eq:floor_additive_gap} establishes the claimed result.
\end{proof}

Thus, \(G_{\mathrm{UB}}\) provides a computable upper bound on the objective loss, and \(G_{\mathrm{UB}}=0\) certifies global optimality.

\textit{Computational complexity.}
Let \(n=|\mathcal{V}|\), \(m=|\mathcal{E}|\), and \(N=|\mathcal{I}|=n-2\). Let \(C_{\mathrm{LP}}\) denote the computational cost of solving the LP relaxation. The componentwise floor rounding of \(\boldsymbol{L}^{\mathrm{LP}}\) requires \(\mathcal{O}(N)\) time.

At each padding iteration, one forward and one reverse topological traversal compute \(\theta_S^{\boldsymbol{L}}\) and \(\theta_T^{\boldsymbol{L}}\) in \(\mathcal{O}(n+m)\) time. Once these values are available, the feasibility and resulting latency of each of the at most \(N\) one-module extensions can be evaluated in constant time. Selecting the feasible extension with the minimum resulting latency requires an additional \(\mathcal{O}(N)\) scan. Since \(N\leq n\), the overall cost of one padding iteration is \(\mathcal{O}(n+m)\).

The padding stage performs at most \(K_{\max}\leq N-1\) successful module additions and may require one additional iteration to detect that no feasible extension remains. Hence, its total computational cost is
\(
\mathcal{O}\!\left((K_{\max}+1)(n+m)\right)=\mathcal{O}\!\left(N(n+m)\right).
\)
Including the LP relaxation and floor rounding, the overall complexity of LPAS is
\(
C_{\mathrm{LPAS}}=C_{\mathrm{LP}}+\mathcal{O}\!\left(N(n+m)\right).
\)

The LP relaxation has polynomial size in the problem instance, and linear programming is solvable in polynomial time. Therefore, LPAS is a polynomial-time synthesis algorithm that returns a feasible architecture with a computable bound on objective suboptimality, without requiring an exact MILP solution.

\section{Model-First Baseline: Layered-Graph Deployment}
\label{lgd}

To compare network-native architecture synthesis with model-first deployment, we construct a baseline optimized over the same physical network. We consider the deployment of a complete predefined DNN over \(\mathcal{G}=(\mathcal{V},\mathcal{E})\), while preserving its operations and data dependencies. Following the layered-graph formulation in \cite{jung2023optimization}, we adapt the state-space construction to the ordered execution-unit profiles and physical-DAG setting considered here, and refer to the resulting baseline as layered-graph deployment (LGD).

\subsection{Complete Reference-DNN Profile}
\label{lgd_profile}

To optimize the deployment of a predefined DNN, we first represent its computation and communication costs at valid partition boundaries. The complete reference DNN is profiled offline as an ordered sequence of deployable execution units \(\mathcal{Q}=(q_1,q_2,\ldots,q_J)\), whose boundaries preserve the operations and data dependencies of the original model. An execution unit may be an individual layer or a composite block containing a residual or multi-branch subgraph.

Each unit \(q_k\) is characterized by its computation workload \(F_k\) and parameter storage \(P_k\). Let \(D_k\) denote the aggregate size of the boundary state after units \(q_1,\ldots,q_k\) have been completed, including any intermediate features required by subsequent units, where \(D_0\) and \(D_J\) denote the raw-input and final-output sizes, respectively. The profile spans the complete reference DNN, from its first model operation through the final classifier, and no execution unit is removed before deployment. Parameter-free operations are retained through their computation and boundary-state contributions.

LGD deploys \(\mathcal{Q}\) over one directed source-to-sink path in \(\mathcal{G}\). Each intermediate node executes a consecutive segment of \(\mathcal{Q}\) or forwards the current boundary state without local DNN computation. The source and sink do not execute reference-DNN units; all units in \(\mathcal{Q}\) are assigned to intermediate compute nodes.

\subsection{Layered-Graph Deployment}
\label{layered_graph}

LGD constructs a layered DAG in which state \((v,k)\) indicates that the first \(k\) execution units have been completed and that the corresponding boundary state is available at physical node \(v\). The initial and terminal states are \((S,0)\) and \((T,J)\), respectively.

For each physical link \((u,v)\in\mathcal{E}\) with \(v\in\mathcal{I}\), a transition \((u,i)\rightarrow(v,j)\), \(0\leq i\leq j\leq J\), is introduced when the segment \(q_{i+1},\ldots,q_j\) satisfies the storage constraint \(\sum_{k=i+1}^{j}P_k\leq M_v\). The transition transmits the boundary state after \(q_i\) to node \(v\) and executes the corresponding segment at \(v\), with weight \(c_{uv}^{ij}=B_{uv}D_i+A_v\sum_{k=i+1}^{j}F_k\). The case \(i=j\) represents pure forwarding. For a physical link \((u,T)\in\mathcal{E}\), the terminal transition \((u,J)\rightarrow(T,J)\) has weight \(c_{uT}^{JJ}=B_{uT}D_J\).

\begin{figure}[t]
    \centering
    \includegraphics[width=\columnwidth]{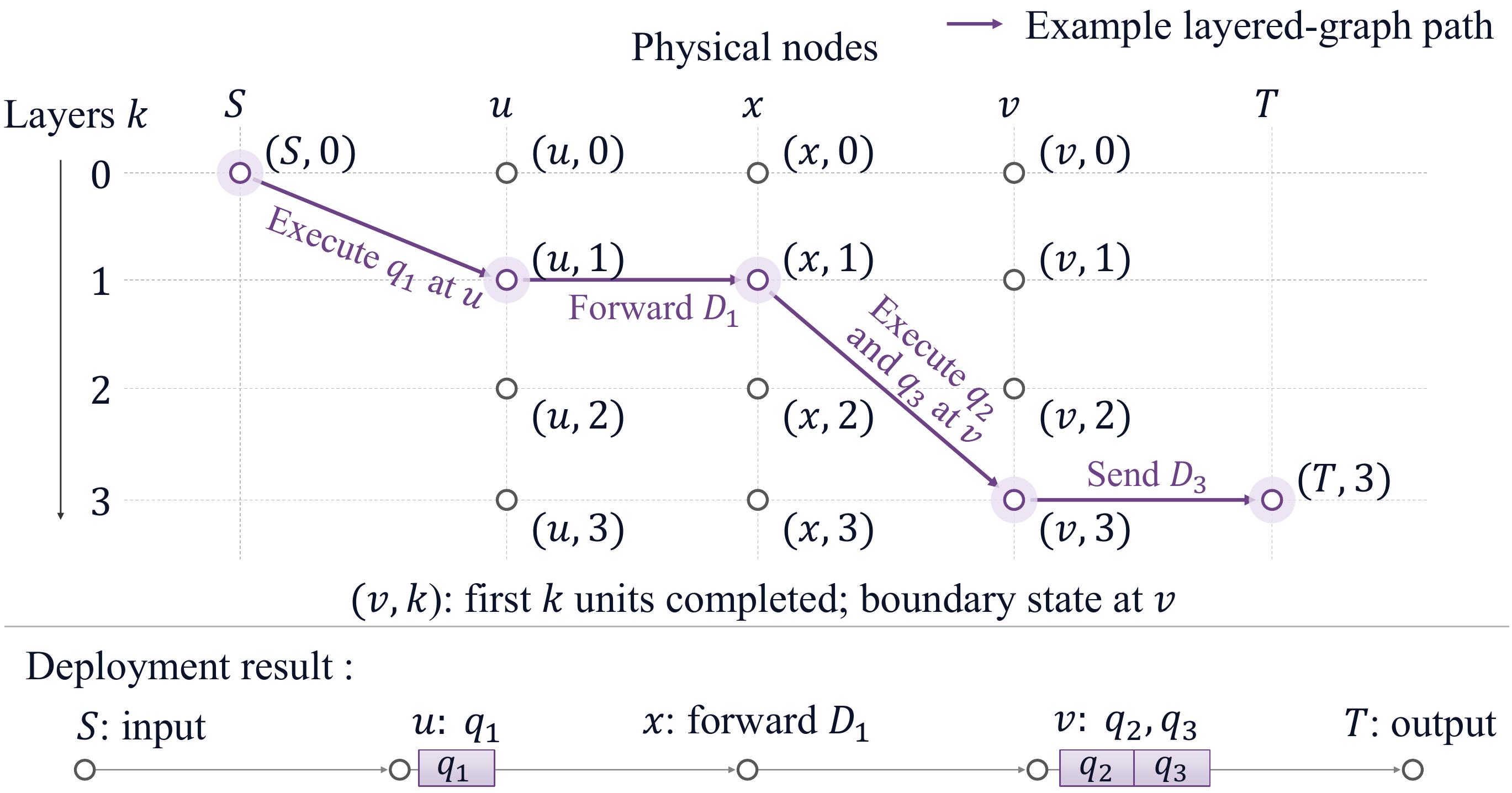}
    \setlength{\abovecaptionskip}{-6pt}
    \caption{Illustrative layered-graph deployment (LGD) with $J=3$ execution units. The highlighted path jointly specifies the physical route and consecutive execution-unit allocation shown below.}
    \label{fig:lgd_layered_graph}
\end{figure}

Fig.~\ref{fig:lgd_layered_graph} illustrates the layered-state construction and how a source-to-sink path jointly specifies the physical route and the consecutive DNN segments executed along that route.

Because the physical network is a DAG, the layered graph is also acyclic. Let \(\mathcal{P}_{\mathrm{LGD}}\) denote the set of directed paths from \((S,0)\) to \((T,J)\). The minimum LGD deployment latency is
\begin{equation}
T_{\mathrm{LGD}}^{\star}
=
\min_{\pi\in\mathcal{P}_{\mathrm{LGD}}}
\sum_{(u,i)\rightarrow(v,j)\in\pi} c_{uv}^{ij}.
\label{eq:lgd_optimal_latency}
\end{equation} 
Every feasible assignment of consecutive execution-unit segments along a single source-to-sink path corresponds to a layered-graph path. Thus, \(T_{\mathrm{LGD}}^\star\) is the minimum latency within this deployment model.
Since the layered graph is acyclic, the shortest path is equivalently obtained by dynamic programming in topological order.

The layered graph contains \(O(|\mathcal{V}|J)\) states and at most \(O(|\mathcal{E}|J^2)\) transitions. Using prefix sums of \(\{F_k\}\) and \(\{P_k\}\), its shortest path can be computed in \(O(|\mathcal{V}|J+|\mathcal{E}|J^2)\) time.

For each reference DNN with a feasible LGD deployment, we use its optimal LGD latency to define the network-execution deadline for the resulting network-native architecture as
\begin{equation}
T_D=T_{\mathrm{LGD}}^{\star}.
\label{eq:lgd_induced_budget}
\end{equation}
Thus, LGD and LPAS operate over the same physical network under a matched network deadline: the model-first baseline optimizes the deployment of the complete predefined DNN within the LGD design space, whereas LPAS synthesizes a network-native architecture subject to the resulting deadline.

\section{Performance Evaluation}
\label{sec:evaluation}

This section evaluates network-native DNN architecture synthesis from both algorithmic and paradigm-level perspectives. We first examine the solution quality, computational cost, and stability of LPAS over physical DAGs of different scales. We then compare the network-native architectures with the model-first LGD baseline under matched network deadlines and examine how the two paradigms utilize the available distributed resources. Finally, we compare classification accuracy on CIFAR-10 and CIFAR-100 across topologies, reference DNNs, and nominal precision settings.

\subsection{Experimental Setup}
\label{experimental_setup}

\begin{figure}[t]
    \centering
    \includegraphics[width=\columnwidth]{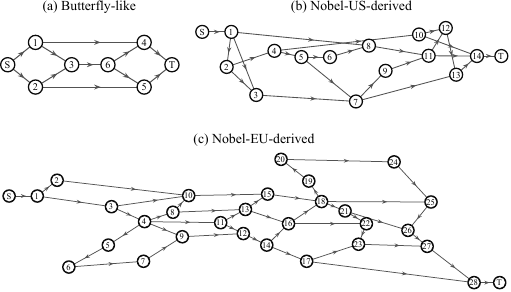}
    \setlength{\abovecaptionskip}{-6pt}
    \caption{Physical DAGs used in the evaluation: (a) butterfly-like DAG with six compute nodes, (b) Nobel-US-derived DAG with 14 compute nodes and 21 benchmark links, and (c) Nobel-EU-derived DAG with 28 compute nodes and 41 benchmark links. \(S\) and \(T\) denote the non-compute input and output interfaces, respectively.}
    \label{fig:evaluation_topologies}
\end{figure}

We consider the three physical DAGs shown in Fig.~\ref{fig:evaluation_topologies}: a butterfly-like DAG with six compute nodes, a Nobel-US-derived DAG with 14 compute nodes, and a Nobel-EU-derived DAG with 28 compute nodes. The latter two are DAG variants derived from the Nobel-US and Nobel-EU benchmark instances in SNDlib, with the directed edges used in our evaluation shown explicitly in Fig.~\ref{fig:evaluation_topologies}. 
For brevity, we refer to the butterfly-like, Nobel-US-derived, and Nobel-EU-derived DAGs as Butterfly, Nobel-US, and Nobel-EU, respectively, throughout the remainder of this paper.

Unless otherwise stated, heterogeneous resources are independently sampled from the same ranges for all three topologies. Node computation throughput is uniformly sampled from \(15\)-\(60\) GFLOP/s, node model-storage capacity from \(3\)-\(22\) MiB, and physical-link rate from \(50\)-\(300\) Mbps. A fixed random seed is used only for reproducibility. The common physical-network and architecture synthesis settings are summarized in Table~\ref{tab:evaluation_configuration}.

Each neural module is instantiated as a dimension-preserving MobileLatentBlock (MLB). At an intermediate compute node with multiple incoming branches, an AttentionFusionBlock (AFB) combines the synchronized parent features. Its computation and storage requirements depend on the number of incoming branches and are included in the corresponding node constraints. At the sink, the fixed output interface aggregates the incoming features before classification. This interface lies outside the modeled network-execution interval. The integer module capacity of each intermediate compute node is determined after accounting for its fusion storage.

\begin{table}[t]
\centering
\caption{Physical Network and Architecture Synthesis Settings}
\label{tab:evaluation_configuration}
\footnotesize
\setlength{\tabcolsep}{3.5pt}
\renewcommand{\arraystretch}{1.05}
\begin{tabular}{@{}ll@{}}
\toprule
Setting & Value \\
\midrule
Node computation throughput & \(15\)-\(60\) GFLOP/s \\
Node model storage & \(3\)-\(22\) MiB \\
Physical-link rate & \(50\)-\(300\) Mbps \\
Common feature tensor & \(64\times32\times32\), FP16 \\
Feature payload & \(0.125\) MiB \\
MLB computation & \(23.156\) MFLOPs \\
MLB parameters & \(69{,}840\) \\
MLB storage & \(0.133\) MiB, FP16 \\
\midrule
AFB computation & \(0.196608k+0.00208k^2\) MFLOPs \\
AFB parameters & \(n_{\mathrm{f}}(k)=1{,}040k^2+33k\) \\
AFB storage & \(2n_{\mathrm{f}}(k)/2^{20}\) MiB, FP16 \\
\bottomrule
\end{tabular}
\par\smallskip
\begin{minipage}{\columnwidth}
\footnotesize
MLB denotes mobile-latent-block and AFB denotes attention fusion block. AFB's costs are given
per node with in-degree \(k\geq2\). Computation is estimated per sample.
\end{minipage}
\end{table}

For the model-first baseline, we use complete profiles of ResNet18 \cite{he2016deep}, DenseNet121 \cite{huang2017densely}, and ResNet50 \cite{he2016deep}, with FP16 representations for parameter storage and intermediate feature transmission. Their profiles contain 46, 132, and 108 deployable execution units, respectively. As described in Section~\ref{lgd}, LGD deploys the complete reference DNN and returns its minimum deployment latency \(T_{\mathrm{LGD}}^\star\).

At the task level, we evaluate network-native architectures synthesized under deadlines induced by ResNet18, ResNet50, and DenseNet121, and compare them with the corresponding model-first baselines on CIFAR-10 and CIFAR-100. For each topology and reference DNN, the network-native models retain the same backbone architecture across the two datasets but use dataset-specific classification heads. All models are randomly initialized and trained from scratch separately on each dataset, with both the backbone and classification head jointly optimized. Each network-native model configuration and each reference DNN configuration is trained three times, and the highest test accuracy across the three runs is reported.

Table~\ref{tab:io_extra_overhead} summarizes the structures, parameter counts, per-image extra computational costs, and FP16 parameter-storage requirements of the fixed input and output interfaces at \(S\) and \(T\).
Both datasets use the same input-reshape architecture: a $3\times3$ convolution followed by batch normalization and SiLU,
mapping RGB images to $64\times32\times32$ features. The output node applies global average pooling (GAP), flattening, dropout, and a fully connected layer with 10 or 100 outputs for CIFAR-10 or CIFAR-100, respectively. These structures are fixed, while their parameters are trained jointly with the backbone separately for each dataset.

The experimental environment, quantization configurations and training hyperparameters are summarized in Table~\ref{tab:experimental_setup}. FP32 denotes the unquantized model configuration, while the W/A notation specifies the bit width of weights and activations, respectively. Quantized models are trained with quantization-aware training (QAT). All models are optimized with SGD with Nesterov momentum with learning-rate warmup followed by linear decay. The peak learning rates for ResNet18, ResNet50 and DenseNet121 are set to 0.025, 0.03 and 0.05, respectively.

The deadline in \eqref{eq:lgd_induced_budget} is matched within the synthesis model using the stated resource profiles. The accuracy experiments do not independently verify end-to-end latency or deadline feasibility at each nominal precision.

LP relaxation and MILP are solved using MATLAB solvers. The MILP solutions serve as references in the quality evaluation of the three topologies, while the MILP solver is also used for runtime comparison in the scalability experiment.

\begin{table}[t]
\centering
\caption{Input and Output Extra Overhead}
\label{tab:io_extra_overhead}
\footnotesize
\setlength{\tabcolsep}{2pt}
\renewcommand{\arraystretch}{1.06}
\begin{tabular*}{\columnwidth}{@{\extracolsep{\fill}}lrrrr@{}}
\toprule
Setting & \multicolumn{2}{c}{CIFAR-10} & \multicolumn{2}{c@{}}{CIFAR-100} \\
\midrule
\multicolumn{5}{@{}l}{\textit{Input at \(S\) (all topologies)}} \\
Structure & \multicolumn{4}{c@{}}{\(3\times3\) Conv (\(3\to64\)) + BN + SiLU} \\
Computation (MFLOPs) & \multicolumn{4}{c@{}}{4.063} \\
Parameters & \multicolumn{4}{c@{}}{1,856} \\
FP16 storage (KiB) & \multicolumn{4}{c@{}}{3.625} \\
\midrule
\multicolumn{5}{@{}l}{\textit{Fusion and classification at \(T\)}} \\
\multicolumn{5}{@{}l}{AFB (if \(k>1\)) \(\to\) GAP \(\to\) Flatten \(\to\) Dropout \(\to \mathrm{FC}_{64\to K}\)} \\
Sink in-degree & \(k=1\) & \(k=2\) & \(k=1\) & \(k=2\) \\
\cmidrule(lr){2-3}\cmidrule(l){4-5}
Computation (MFLOPs) & 0.132 & 0.534 & 0.144 & 0.546 \\
Parameters & 650 & 4,876 & 6,500 & 10,726 \\
FP16 storage (KiB) & 1.270 & 9.523 & 12.695 & 20.949 \\
\bottomrule
\end{tabular*}
\par\smallskip
\begin{minipage}{\columnwidth}
\footnotesize
T node maps \(k\) incoming features to \(K\) class scores.
Nobel-US and Nobel-EU each supply one feature (\(k=1\)).
On Butterfly (\(k=2\)), the head fuses the features from \(N_4\) and
\(N_5\) before pooling and classification. Totals cover the complete
head. Computation is reported per sample.
\end{minipage}
\end{table}

\begin{table*}[t]
    \centering
    \caption{Implementation and training settings.}
    \label{tab:experimental_setup}
    \small
    \renewcommand{\arraystretch}{1.25}
    \setlength{\tabcolsep}{10pt}

    \renewcommand{\tabularxcolumn}[1]{m{#1}}

    \begin{tabularx}{\textwidth}{
        @{} *{3}{>{\centering\arraybackslash}X} @{}
    }
        \toprule
        \multicolumn{1}{c}{\textbf{Hardware and Software}}
        & \multicolumn{1}{c}{\textbf{Quantization Settings}}
        & \multicolumn{1}{c}{\textbf{Hyperparameters}} \\
        \midrule
        \textbf{GPU:} RTX6000D \par
        \textbf{VRAM:} 84~GB \par
        \textbf{PyTorch:} 2.7.1 \par
        \textbf{Torchvision:} 0.22.1 \par

        &

        \textbf{Method:} Full precision (FP32); QAT (W8A8/W4A4) \par
        \textbf{Network-native:} FP32 / W8A8 / W4A4 \par
        \textbf{Baselines:} FP32 / W8A8 / W4A4 \par

        &

        \textbf{Optimizer:} SGD with Nesterov \par
        \textbf{Batch size:} 64 \par
        \textbf{Epochs:} 5 (warm-up)+ 45 (training)\par
        \textbf{Peak LR:} $\{0.025,\,0.03,\,0.05\}$ \par
        \textbf{LR schedule:} Warmup + linear decay \par
        \\
        \bottomrule
    \end{tabularx}
\end{table*}

\subsection{LPAS Algorithm Evaluation}
\label{subsec:lpas_evaluation}

We first isolate the synthesis algorithm from task-level training and examine whether the LP-floor-greedy procedure consistently produces high-quality integer solutions with low computational overhead.

\subsubsection{Solution Quality and Runtime}
\label{subsubsec:lpas_quality}

We first evaluate LPAS on the three physical DAGs in Fig.~\ref{fig:evaluation_topologies}. For each topology, we generate 50 independent heterogeneous resource realizations using the ranges in Table~\ref{tab:evaluation_configuration}. For a realization with node-wise module-capacity vector \(\boldsymbol{U}\), let \(T_{\min}=T(\boldsymbol{0})\) and \(T_{\max}=T(\boldsymbol{U})\). To cover latency regimes from tight to loose while excluding the two endpoint cases, we set
\(
T_D(\lambda)=T_{\min}+\lambda\bigl(T_{\max}-T_{\min}\bigr),
\)
where \(\lambda\in\{0.05,0.15,\ldots,0.95\}\). This yields 500 resource-deadline instances per topology.

For each instance, let \(Z_{\mathrm{LP}}^\star\) denote the LP optimum and \(Z_{\mathrm{UB}}=\lfloor Z_{\mathrm{LP}}^\star\rfloor\) the corresponding integer upper bound. We evaluate the componentwise floor solution \(Z_0\) and the final LPAS solution \(Z_{\mathrm{LPAS}}\) through their upper-bound gaps, \(K_{\max}=Z_{\mathrm{UB}}-Z_0\) and \(G_{\mathrm{UB}}=Z_{\mathrm{UB}}-Z_{\mathrm{LPAS}}\), respectively. Table~\ref{tab:lpas_physical_dags} reports the mean upper-bound gaps and runtimes across the three physical DAGs.

\begin{table}[t]
\centering
\caption{Algorithm performance on the three physical DAGs.}
\label{tab:lpas_physical_dags}
\footnotesize
\renewcommand{\arraystretch}{1.25}
\setlength{\tabcolsep}{5pt}
\begin{tabular*}{0.8\columnwidth}{@{\extracolsep{\fill}}lcccc@{}}
\hline
& \multicolumn{2}{c}{Gap (modules)} & \multicolumn{2}{c}{Runtime (ms)} \\
\cline{2-3}\cline{4-5}
Physical DAG & $K_{\max}$ & $G_{\mathrm{UB}}$ & LPAS & MILP \\
\hline
Butterfly & 0.510 & 0.250 & 4.52 & 5.30 \\
Nobel-US & 0.510 & 0.218 & 4.53 & 6.28 \\
Nobel-EU & 0.888 & 0.420 & 5.08 & 18.89 \\
\hline
\end{tabular*}
\end{table}

As shown in Table~\ref{tab:lpas_physical_dags}, greedy padding reduces the mean upper-bound gap by approximately half across all three DAGs. LPAS also maintains low runtime, ranging from 4.52 to 5.08 ms, with a modest increase as the topology grows. These results demonstrate that LPAS produces high-quality feasible allocations with little computational overhead on the physical networks considered in the task-level evaluation.

We further evaluate LPAS on larger randomly generated physical networks to examine its solution quality and computational scalability. Specifically, we generate single-source single-sink DAGs with 30 to 70 compute nodes in increments of 10. For each network size, we consider 50 independent topology and resource realizations using the same resource ranges as in Table~\ref{tab:evaluation_configuration}. The random DAGs are generated by distributing compute nodes across ordered layers and establishing forward connections to ensure source-to-sink reachability. Additional forward edges are randomly added until the total number of links is approximately 2.5 times the number of compute nodes. For solution quality, we select \(N\in\{30,50,70\}\) and vary \(\lambda\) from 0.1 to 0.9 in increments of 0.1, reusing the same realizations across deadlines. We measure the relative upper-bound gap of LPAS as \(100(Z_{\mathrm{UB}}-Z_{\mathrm{LPAS}})/Z_{\mathrm{UB}}\). Since \(Z_{\mathrm{UB}}\) is an upper bound on the integer optimum, this metric also upper-bounds the relative optimality gap. For runtime, we fix \(\lambda=0.5\) and compare LPAS with direct MILP optimization across all five network sizes. For the scalability experiment, the MILP solver uses a 10-second time limit and an absolute optimality-gap tolerance of 0.99 modules.

\begin{figure}[t]
\centering
\includegraphics[width=\linewidth]{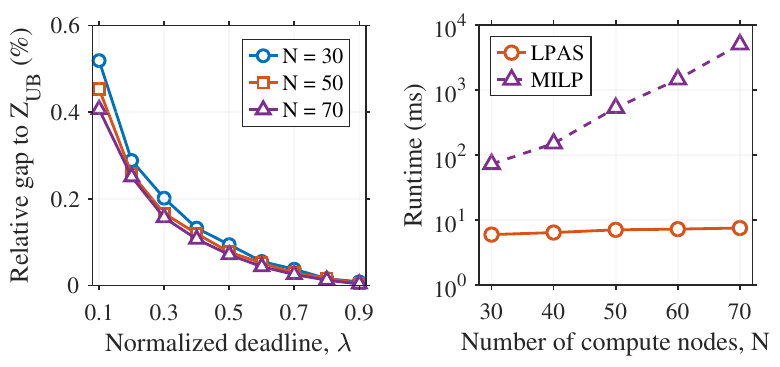}
\setlength{\abovecaptionskip}{-8pt}
\caption{Solution quality and runtime of LPAS on random DAGs. Left: mean relative gap to the LP-derived integer upper bound \(Z_{\mathrm{UB}}\) versus normalized deadline \(\lambda\) for three network sizes. Right: median runtimes of LPAS and MILP versus the number of compute nodes at \(\lambda=0.5\), shown on a logarithmic scale. Each point summarizes 50 independent topology and resource realizations.}
\label{fig:algorithm_performance}
\end{figure}

The left panel of Fig.~\ref{fig:algorithm_performance} shows that LPAS maintains a small relative upper-bound gap across different network sizes and latency constraints. The mean gap remains below 0.6\% for all tested configurations and decreases toward zero as the deadline is relaxed. The three curves also exhibit similar behavior, indicating that LPAS consistently achieves high-quality solutions across the considered network sizes.

The right panel demonstrates the computational efficiency of LPAS as the network size increases. Its median runtime remains below 10 ms across all tested sizes, whereas direct MILP optimization requires increasingly more time, rising from tens of milliseconds to several seconds. These results show that LPAS maintains a small relative gap to the LP-derived upper bound while requiring substantially less computation time than direct MILP optimization on the tested larger DAGs.

\subsubsection{Perturbation Stability}
\label{subsubsec:lpas_perturbation}

We next examine whether bounded variations in the physical-network parameters cause abrupt changes in the synthesized architecture. For each topology, we reuse the 50 nominal resource realizations and select the representative interior operating point \(\lambda=0.5\). The nominal deadline is computed before perturbation and then held fixed.

For each nominal realization, independent perturbation directions \(\xi_v,\xi_{uv}\sim\mathcal{U}[-1,1]\) are drawn once and scaled according to \(A_v^{(\rho)} = A_v(1+\rho\xi_v)\) and \((B_{uv}D)^{(\rho)} = B_{uv}D(1+\rho\xi_{uv})\), where \(\rho\in\{0,0.02,\ldots,0.10\}\). The same perturbation direction is retained across all values of \(\rho\) for a given realization, yielding a paired perturbation sweep. The physical topology, node-storage capacities, and nominal deadline remain.

Let \(\widehat{\boldsymbol{L}}^{(0)}\) and \(\widehat{\boldsymbol{L}}^{(\rho)}\) denote the nominal and perturbed allocations, respectively, and define \(Z^{(\rho)}=\sum_{v\in\mathcal{I}}\widehat{L}_v^{(\rho)}\). 
We measure the relative change in architecture size as $\Delta_Z^{(\rho)}=100|Z^{(\rho)}-Z^{(0)}|/Z^{(0)}$. The allocation-share deviation is defined as $\Delta_{\mathrm{alloc}}^{(\rho)}=50\|\hat{\boldsymbol{L}}^{(\rho)}/Z^{(\rho)}-\hat{\boldsymbol{L}}^{(0)}/Z^{(0)}\|_1$.
The latter is the total-variation distance between the normalized node-wise allocation shares, expressed as a percentage, and therefore captures module redistribution independently of changes in the total architecture size.

\begin{figure}[t]
    \centering
    \includegraphics[width=\columnwidth]{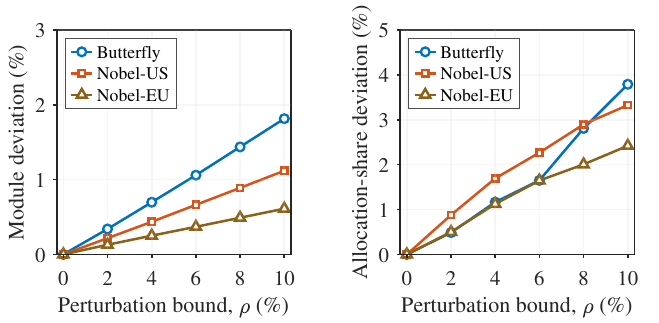}
    \setlength{\abovecaptionskip}{-6pt}
    \caption{Stability of LPAS under bounded computation and communication perturbations. The plots report the mean module-count deviation and mean allocation-share deviation over 50 paired realizations per topology at \(\lambda=0.5\).}
    \label{fig:lpas_perturbation}
\end{figure}

As shown in Fig.~\ref{fig:lpas_perturbation}, both deviations increase gradually with the perturbation bound rather than exhibiting abrupt changes. Even under the largest \(10\%\) perturbation, the mean total-module deviation remains below \(2\%\), while the mean allocation-share deviation remains below \(5\%\) for all three topologies. The larger change in allocation shares indicates that parameter variations primarily induce modest redistribution of modules among compute nodes while leaving the overall architecture size comparatively stable. The consistent behavior across all three DAGs indicates that LPAS is not excessively sensitive to moderate perturbations in the computation and communication parameters.

\subsection{Network-Native Synthesis Versus Model-First Deployment}
\label{subsec:paradigm_comparison}

We next compare network-native architecture synthesis with model-first deployment under matched network deadlines.
 
For the paradigm-level comparison, we use one fixed physical-resource configuration within the ranges in Table~\ref{tab:evaluation_configuration} for each topology and apply it consistently to all three reference DNNs. No model-specific resource retuning is performed. For each reference DNN \(r\), LGD first determines its minimum deployment latency \(T_{\mathrm{LGD},r}^{\star}\), after which LPAS synthesizes a network-native architecture over the same physical network with \(T_{D,r}=T_{\mathrm{LGD},r}^{\star}\). Thus, each comparison uses identical physical-network resources and a matched network deadline, while differing in whether the DNN architecture is fixed before deployment or synthesized according to the physical network.

\subsubsection{Resource Utilization}
\label{subsubsec:resource_utilization}

We use two direct metrics to characterize how the two paradigms exploit the distributed resources. For \(X\in\{\mathrm{LGD},\mathrm{NNA}\}\), where \(\mathrm{NNA}\) denotes the network-native architecture, let \(\mathcal{I}_{\mathrm{act}}^{(X)}\) denote the compute nodes hosting at least one DNN execution unit under LGD or at least one synthesized module under the network-native synthesis scheme, and let \(P_{\mathrm{dep}}^{(X)}\) denote the total inference-model storage deployed across the compute nodes. We define \(R_{\mathrm{act}}^{(X)}=|\mathcal{I}_{\mathrm{act}}^{(X)}|/|\mathcal{I}|\) and \(R_{\mathrm{stor}}^{(X)}=P_{\mathrm{dep}}^{(X)}/\sum_{v\in\mathcal{I}}M_v\).
The active compute-node ratio measures how broadly DNN execution units or synthesized modules are distributed over the physical network, whereas the aggregate model-storage utilization measures the fraction of the total compute-node storage occupied by the deployed inference model.

For LGD, \(P_{\mathrm{dep}}^{(\mathrm{LGD})}\) includes the complete reference-DNN parameters deployed on intermediate compute nodes. For the network-native architecture, \(P_{\mathrm{dep}}^{(NNA)}\) includes all MLBs and AFBs synthesized on intermediate nodes. The lightweight input and output interfaces at \(S\) and \(T\) lie outside the modeled network-execution interval and are therefore excluded from the intermediate-node storage accounting.

\begin{figure}[t]
    \centering
    \includegraphics[width=\columnwidth]{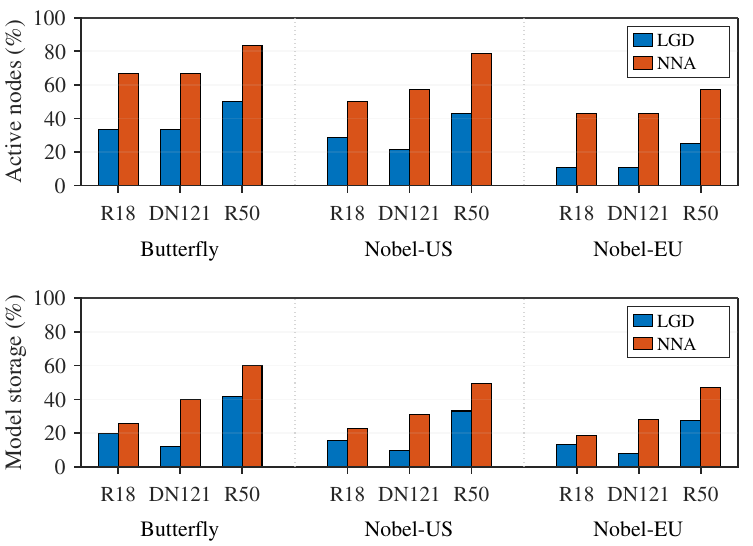}
    \setlength{\abovecaptionskip}{-6pt}
    \caption{Resource utilization of LGD and network-native architectures under matched deadlines. Top: active compute-node ratio. Bottom: aggregate model-storage utilization. The three groups from left to right correspond to Butterfly, Nobel-US, and Nobel-EU. R18, DN121, and R50 denote ResNet18, DenseNet121, and ResNet50, respectively.}
    \label{fig:resource_utilization}
\end{figure}

Figure~\ref{fig:resource_utilization} reveals a consistent structural difference between the two paradigms. The network-native architectures activate a larger fraction of compute nodes in every topology-model pair, whereas LGD performs DNN computation only along its selected deployment path. The difference becomes more pronounced on the larger DAGs, where the network-native architectures can exploit additional physical branches that are unavailable to a single-path model-first deployment.

The network-native architectures also occupy a larger fraction of the aggregate model-storage capacity across the evaluated cases. This observation should not by itself be interpreted as higher resource efficiency or better task performance. Rather, it indicates that, under the same LGD-induced network deadline, network-native architecture synthesis can convert a broader portion of the distributed storage resources into trainable model capacity, whereas LGD remains constrained by its predefined architecture and selected deployment path. Whether this additional architectural capacity translates into improved task performance is examined in the subsequent task-level comparison.

\subsubsection{Task Performance}
\label{subsubsec:task_performance}
\begin{figure}[t]
    \centering
    \includegraphics[width=\linewidth]{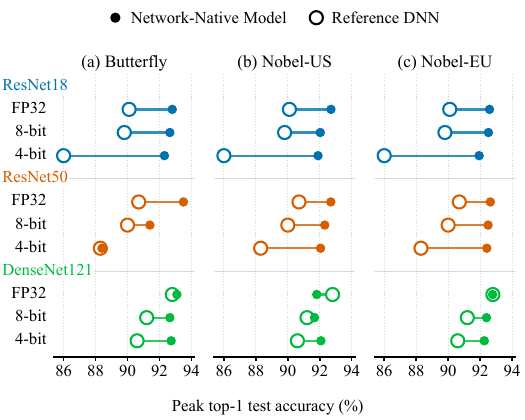}
    \setlength{\abovecaptionskip}{-8pt}
    \caption{CIFAR-10 best test accuracy on (a) Butterfly, (b) Nobel-US, and (c) Nobel-EU. Each panel contains nine reference DNN pairs. Filled circles represent the network-native models and hollow circles represent the reference DNN deployed by LGD. Colors distinguish ResNet18, ResNet50, and DenseNet121. Each segment joins a matched pair of reported values, not an uncertainty interval. The horizontal scale is shared across panels.}
    \label{fig:task_accuracy_cifar10}
\end{figure}

\begin{figure}[t]
    \centering
    \includegraphics[width=\linewidth]{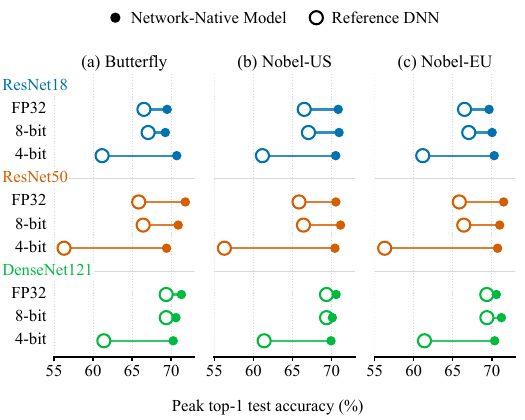}
    \setlength{\abovecaptionskip}{-8pt}
    \caption{CIFAR-100 best test accuracy, with the same topology panels, model colors, nominal precision rows, and marker convention as Fig.~\ref{fig:task_accuracy_cifar10}. The horizontal scale is shared within this figure, but differs from CIFAR-10.}
    \label{fig:task_accuracy_cifar100}
\end{figure}

We compare network-native models with complete reference DNNs deployed by LGD. For each topology, LGD finds the minimum deployment latency of each reference DNN which is the network-level inference budget of corresponding network-native models. Network-native then synthesizes its architecture from the topology under this budget. The left-side labels in the figures~\ref{fig:task_accuracy_cifar10} and \ref{fig:task_accuracy_cifar100} identify the DNNs used to set these budgets.

Figures~\ref{fig:task_accuracy_cifar10} and \ref{fig:task_accuracy_cifar100} report peak top-1 test accuracies on CIFAR-10 and CIFAR-100. Each topology panel covers three reference DNNs at three precision settings. LGD preserves the reference DNNs during deployment. For a given dataset and precision, each reference DNN has one baseline accuracy. We use this value across all three topologies.
The two figures contain 54 network-native models accuracy values and 18 distinct baseline values.
For each dataset and precision setting, Table~\ref{tab:task_accuracy_means} averages network-native models accuracy over the nine topology and precision settings. The baseline mean averages the three reference DNNs.

\noindent\textbf{CIFAR-10.}
The network-native models achieve 88.40-93.50\% accuracy and exceed the baseline in 25 of 27 configurations. The paired gains range from $-0.99$ to $6.31$ percentage points, with a mean gain of $2.31$ percentage points across the 27 evaluated configurations. The highest accuracy among the network-native models is 93.50\%, attained by the network-native model synthesized under the ResNet50-matched deadline on Butterfly, compared with 90.70\% for its FP32 baseline. All nine network-native model configurations with ResNet18-matched deadlines outperform their baselines, with accuracies of 91.89-92.80\%. Overall, the network-native models achieve accuracy gains across all three reference DNNs in most configurations, while remaining within one percentage point of the baseline in the two exceptions.

\noindent\textbf{CIFAR-100.}
The network-native models achieve 69.25-71.81\% accuracy and exceeds the baseline in all 27 configurations. Gains range from $0.74$ to $14.45$ percentage points and have a descriptive mean of $5.71$ percentage points, a larger mean gap than on CIFAR-10. At FP32, the network-native models synthesized under ResNet50-matched deadlines achieve 70.56\%, 71.51\%, and 71.81\% on Nobel-US, Nobel-EU, and Butterfly, respectively, versus 65.84\% for the baseline. The corresponding gains of 4.72, 5.67, and 5.97 percentage points show that the advantage also occurs in the nominal FP32 setting. The FP32 gains are smaller for DenseNet121 (1.22-1.95 percentage points) and ResNet18 (2.97-4.35 percentage points), demonstrating that the improvement depends on the reference-model family.

\begin{table}[t]
\centering
\caption{Mean Classification Accuracy and Gain over the Model-First Baseline}
\label{tab:task_accuracy_means}
\footnotesize
\setlength{\tabcolsep}{3pt}
\renewcommand{\arraystretch}{1.1}
\begin{tabular*}{0.9\columnwidth}{@{\extracolsep{\fill}}llccc@{}}
\toprule
Dataset & \shortstack{Nominal\\precision} & \shortstack{Baseline\\(\%)} & \shortstack{Network-native\\(\%)} & \shortstack{Gain\\(pp)} \\
\midrule
CIFAR-10 & FP32  & 91.20 & 92.73 & +1.53 \\
         & 8-bit & 90.33 & 92.24 & +1.91 \\
         & 4-bit & 88.30 & 91.79 & +3.49 \\
\midrule
CIFAR-100 & FP32  & 67.24 & 70.71 & +3.47 \\
          & 8-bit & 67.61 & 70.59 & +2.98 \\
          & 4-bit & 59.61 & 70.30 & +10.69 \\
\bottomrule
\end{tabular*}
\par\smallskip
\begin{minipage}{\columnwidth}
Means summarize reported best test accuracies. The network-native model mean gives equal weight to nine topology-model configurations per row; the baseline averages three reference models, each reused across topologies. Gain is the network-native model mean minus the baseline mean in percentage points (pp), calculated from the displayed rounded means. These are configuration averages, not repeated-seed estimates.
\end{minipage}
\end{table}

\noindent\textit{Precision and topology dependence.}
Table~\ref{tab:task_accuracy_means} shows positive descriptive mean gains at all three nominal precisions. On CIFAR-10, the reported mean gains are 1.53 percentage points at FP32, 1.91 at 8-bit, and 3.49 at 4-bit. On CIFAR-100, the corresponding gains are 3.47, 2.98, and 10.69 percentage points. The largest reported mean gap occurs at nominal 4-bit on CIFAR-100, where the baseline mean is 59.61\% and the network-native model mean is 70.30\%. Relative to their FP32 means, these represent decreases of 7.63 and 0.41 percentage points, respectively. For the network-native models on CIFAR-10, the precision decreases 0.94 percentage points from FP32 to 4-bit. These differences are calculated from the rounded table entries. 
Individual configurations do not always follow the aggregate trend. These precision comparisons evaluate fixed architectures rather than architectures independently synthesized for each precision setting.

The topology ranking also changes with the reference model and precision. For network-native models synthesized under ResNet50-matched deadlines on CIFAR-10, Butterfly has the highest FP32 accuracy, whereas Nobel-EU has the highest 4-bit accuracy. On CIFAR-100 at FP32, Nobel-US leads for ResNet18, while Butterfly leads for ResNet50 and DenseNet121. Thus, the largest physical DAG does not consistently yield the highest accuracy.

\section{{Conclusion}}
\label{sec:conclusion}

This paper developed a network-native DNN architecture synthesis framework and an efficient algorithm for constructing distributed neural architectures under physical-network constraints.
The results highlight an important insight: the physical network can serve not only as the infrastructure for executing neural computation, but also as a resource for shaping the neural architecture itself. 
By exploiting distributed computation and storage across the network, architecture synthesis creates opportunities to utilize resources that may remain underused when neural dependencies are fixed in advance. The observed improvements in classification accuracy further suggest that this architectural flexibility can yield benefits beyond resource utilization, extending to task-level performance. 
Overall, these findings demonstrate the potential of treating neural architecture as a network-dependent design variable, rather than a fixed input to deployment optimization. 
This perspective opens a new avenue for distributed intelligence, where neural architectures and the underlying computing networks are designed as an integrated system.

\bibliographystyle{IEEEtran}
\bibliography{ref}

\end{document}